\documentstyle[preprint,aps]{revtex}
\begin{document}
\title{Thermal behavior of Quantum Cellular Automaton wires}
\author{C. Ungarelli\footnote{Current address: School of Mathematical Sciences,
University of 
Portsmouth, Mercantile House, Hampshire Terrace, Portsmouth P01 2EG, UK},
S. Francaviglia, M. Macucci, G. Iannaccone}
\address{Dipartimento di Ingegneria dell'Informazione,
Universit\`a di Pisa\\
Via Diotisalvi, 2, I-56126 Pisa, Italy}
\date{\today}
\maketitle
\begin{abstract}
We investigate the effect of a finite temperature on the behavior of 
logic circuits based on the principle of Quantum Cellular Automata (QCA) and
of ground state computation. In particular, we focus on the error 
probability for a wire of QCA cells that propagates a logic state. 
A numerical model and an analytical, more approximate, model are presented 
for the evaluation of the partition function of such a system and, consequently,
of the desired probabilities. We compare the results of the two models, 
assessing
the limits of validity of the analytical approach, and provide estimates for 
the maximum operating temperature.
\end{abstract}
\narrowtext
\section{Introduction}
In recent years a new paradigm for computation has been proposed 
by Craig Lent and coworkers\cite{LentAPL}, based on the concept
of Quantum Cellular Automata (QCA). Such a concept, although extremely
difficult to implement from a technological point of view\cite{JAPnos}, has 
several
interesting features that make it worth pursuing. The basic building
block is made up of a single cell, containing two electrons that 
can be localized in four different areas or ``dots,'' located at the 
vertices of a square, as shown in each of the cells represented in 
Fig.~\ref{eins}(a). Coulomb repulsion forces the two electrons to occupy dots 
that are aligned along one of the diagonals, and each of the two possible 
alignments is associated with a logic state. By placing cells next to each 
other, 
a wire can be formed (binary wire), along which polarization enforced at one 
end will
propagate\cite{LentJAP}, as a consequence of the system of charges relaxing
down to the ground state. We can see this also as the logic state of the 
first cell propagating down the chain until it reaches the last cell. It has
been shown that, by 
properly assembling two-dimensional arrays
of cells, it is possible to implement any combinatorial logic 
function\cite{LentJAP}. The basic principle of operation of such circuits is 
therefore 
the relaxation of the system to the ground state, thus leading to the 
often used expression ``ground-state computation''.

Even in the case of perfectly symmetric
and identical cells, the configuration of the QCA circuit may depart 
from the ground state as a consequence of thermal excitations. If the 
energy separation between the ground state and the first few excited states
is small, their occupancy will be nonnegligible even at low temperatures,
and the logic output may be corrupted. A complete understanding of 
the behavior of QCA arrays as a function of temperature is thus
essential for any practical application of the QCA concept. The 
problem of errors due to finite temperature operation was first 
addressed by Lent\cite{Lententr}, on the basis of entropy considerations.
Our approach consists in a detailed study of thermal statistics for 
QCA arrays, retrieving the results of Ref.\cite{Lententr} as a 
special case, and allowing treatment of cells with more than just two 
states. We have developed both a 
numerical model, which enables us to study relatively short
chains made up of six-state cells in full detail, and an analytical model, 
which can be used for arbitrarily long chains of two-state cells. In both
cases, we have considered a semiclassical approximation, and computed the 
probability of the system being in the ground
state and that of presenting the correct logic output, i.e. of having the
last cell of the chain in the expected logic state. Since only one
configuration corresponds to the ground state, while several different 
configurations
are characterized by the correct logic output, the probability of having
the correct output is always larger than that of being exactly in the ground 
state.

In Sec.~\ref{enesec} we present the cell model we have considered for 
both approaches and the
semi-classical approximation that we have chosen to adopt. We also discuss the
structure of the energy spectrum for the excited states of a chain 
of cells. In Sec.~\ref{partisec} we present the procedure that has been 
followed for the calculation of the partition function with the numerical
method and the associated results for the probabilities of correct operation
as a function of temperature. The analytical model is described in 
Sec.~\ref{gebbo}, together with the associated results and a comparison 
with those from the numerical model.

\section{Model}\label{enesec}
Our approach is semi-classical insofar as electrons are treated as classical 
particles, with the only additional property that they can tunnel between 
dots belonging to the same cell. This is a reasonable approximation if 
the tunneling matrix elements between the dots of a cell are small 
enough to strongly localize the electrons, which therefore behave as 
well defined particles. 

Our model chains are characterized by two geometrical parameters: $d$ is 
the distance between neighboring cell centers, $a$ is the distance between
two dots in a cell. We represent the {\sl driver} cell, i.e. the cell
whose polarization state is externally enforced, with bold lines and 
indicating only the electron positions (see Fig.~\ref{eins}(a)); {\sl driven} 
cells are represented with solid lines and each dot is indicated with a solid
circle if occupied or with an empty circle otherwise. 
Within each cell we consider 
a uniformly distributed ($e/2$ per dot, where $e$ is the electron charge)
positive background charge, which makes each cell overall neutral and 
prevents anomalous behaviors in the nearby cells, due to the uncompensated
monopole component of the electrostatic field. In particular, the repulsive
action of the uncompensated electrons in a driver cell can ``push'' the 
electrons in the nearby driven cell away, thus leading to the formation of
an unwanted state in which electrons are aligned along the side further from 
the driver cell. 

In our calculations we have considered the GaAs/AlGaAs 
material system and assumed a uniform relative permittivity of 12.9: 
this is a reasonable approximation, since the 
permittivity of AlGaAs does not differ significantly from that of the GaAs
layer, where the electrons are confined. For this study we have neglected, for 
the sake of simplicity and of generality, the effects of the semiconductor-air 
interface and of the metal gates defining the dots, whose rigorous treatment 
would have required considering a specific layout\cite{thin}.

For silicon-on-insulator QCA cells\cite{tubopap}, materials with quite  
different permittivities come into play: silicon, silicon oxide and air, 
but reasonable estimates could be obtained by repeating our 
calculations with a relative permittivity corresponding to that of silicon
oxide, since most of the electric field lines are
confined in 
the oxide 
region embedding the silicon dots. Moreover, estimates of the 
performance obtained with this approximation would be conservative, since part
of the field lines are actually in the air over the device, whose relative
permittivity is unitary, thus leading to a
stronger electrostatic interaction and therefore to a reduced importance of
thermal fluctuations.

As we have already stated, the two minimum energy configurations of a cell 
are those with the electrons aligned along one of the diagonals, since 
these correspond to the maximum separation between the electrons. However, 
other 
configurations are also possible, and, depending on intercell spacing,
they can appear in the first few excited states of a binary wire. 
We consider all of the six configurations that can be assumed by two electrons 
in four dots, excluding only those with both electrons in the same dot, which
correspond to too large an energy.

We define the two lowest energy configurations (those with the electrons 
along the diagonals) state 1 and state 0 as indicated in Fig.~\ref{eins}(b), 
while the corresponding polarization values are $1$ and $-1$, respectively. 
Polarization 
values are defined\cite{4N+2} as 
\begin{equation}
P={{Q_1+Q_3-Q_2-Q_4}\over{2}},
\end{equation}
where $Q_i$ is the charge in the $i$-th dot, with the first dot being at the 
top right and the others numbered counterclockwise. Configurations with the
two electrons along one of the four sides of the cell have higher
energies, as stated before, and do not correspond to a well defined logic 
state. For this reason, we define them as $X$ states.

The energy is computed as the electrostatic energy of a classical system
of charges\cite{Jackson}:
\begin{equation}
E={\sum_{i\neq j}{\frac{q_{i}q_{j}}{4 \pi \epsilon _{0}
\epsilon_{r} r_{ij}}}}
\end{equation}
Since in our model the total charge in each dot is either the background charge
(empty dot) or the algebraic sum of the background charge and the charge of
an electron, it can take on only two values: $+e/2$ or $-e/2$, which implies 
that
\begin{equation}
q_{i}q_{j}=\frac{1}{4}e^2 {\rm sgn}(q_{i}q_{j}).
\end{equation}
If we write the interelectronic distance $r_{ij}$ in terms of the 
ratio $R=d/a$ and of the
configuration, the energy of a binary wire can be written as
\begin{equation}
E={\frac{e^{2}}{4a}\frac{1}{4\pi\epsilon_{0}\epsilon_{r}}\sum_{i\neq j}
{\frac{s_{ij}}{\sqrt{\left( n_{ij} R + l_{ij} \right)^{2} + m_{ij}^{2}}}}},
\label{eq1}
\end{equation}
where $n_{ij} \in \left\{ 0,\dots,N_{cell}-1 \right\}$ is the number of cells 
between the cell containing dot $i$ and the cell containing dot $j$, 
$s_{ij}\in \left\{ -1,1 \right\}$ indicates the sign of $q_{i}q_{j}$,
$l_{ij}\in \left\{-1,0,1\right\}$ and
$m_{ij}\in \left\{0,1\right\}$, indicate the position of dots $i$ and $j$
inside the corresponding cells. In particular, $l_{ij}$ is 
equal to 0 if both dots $i$ and $j$ are on the left side or on the right side 
of the cell, to -1 if dot $i$ is on the right side and dot $j$ is on the left 
side and to 1 if dot $i$ is on the left side and dot $j$ is on the right one. 
Furthermore, $m_{ij}$ is equal to 0 if both dots $i$ and $j$ are on the top or 
on the bottom of a cell, to 1 if one dot is on the top and the other is on the 
bottom. 

We have considered a binary wire made up of six cells (one of which is a 
driver cell in a fixed polarization state) with size $a=40$~nm and 
computed the 
energy values corresponding to all possible $6^5$ configurations. The
values thus obtained have been ordered with the purpose of studying the 
energy spectra for different parameter choices. Let us define $R=d/a$. If 
$R\gg 1$, i.e. $d\gg a$,
the interaction between neighboring cells is substantially due to just the 
dipole component, and a discrete spectrum is observed already for $R=2.5$
(see Fig.~\ref{zwei}), with clear steps: 
the ground state corresponds to configurations with all cells in the
same logic state: either all 1 or 0; the first excited state, for $R=2.5$,
includes configurations with one ``kink,'' i.e. with
one cell flipped with respect to the rest of the chain. Higher steps 
correspond to a larger number of kinks. Energy values are expressed with 
reference to the ground state energy and in kelvin, i.e. as the result of the 
division of the actual energies in joule by the Boltzmann constant.

If $R$ is decreased, the interaction between neighboring cells is
incremented and made more complex, so that $X$ states do appear, as
shown in Fig.~\ref{zwei} for $R=1.75$. The various plateaus start merging and a 
continuous spectrum is approached.  
In particular, if we decrease $R$ while keeping $a$ constant, and thereby
reducing the separation between neighboring cells, the difference between 
the energy of the ground state and that of the first excited state is expected
to increase as $1/R$, due to the increased electrostatic interaction. 
However, this
is true only down to a threshold value of $R$, below which the splitting
between the first excited state and the ground state starts decreasing, 
as shown in Fig.~\ref{drei}, where the energy split is plotted as a function 
of $R$ for a cell size $a$ of 40~nm, for a wire with 2 (dotted line), 
3 (dashed line),
and 6 (solid line) cells. This sudden change of behavior can be 
understood on the basis of the previously discussed results: below the 
threshold value for $R$, the configuration for the first excited state contains
a cell in the $X$ state, thereby disrupting the operation of the wire and 
lowering the splitting between the first excited state and the ground state.

In the inset of Fig.~\ref{drei} we report the dependence of the splitting 
between the
two lowest energy states on the number of cells. These results are for 
$R=2.5$, i.e. for a condition in which no $X$ state appears. Once the number of
cells is larger than a few units, the splitting quickly saturates to a constant
value. This is easily understood if we consider that the first excited state 
is characterized by the cell at the end of the wire being polarized opposite 
to the others: the strength of the electrostatic interaction drops quickly
along the chain, and hence no significant change is determined by the 
addition of cells beyond the first five or six. 
The energy splitting has been computed for a cell size $a=40$~nm and,  
as can be deduced from Eq.(\ref{eq1}), is inversely proportional to $a$. 
It can thus be increased by scaling down cell dimensions.

\section{Numerical results for the thermal behavior}\label{partisec} 

In order to compute the probabilities, at a finite temperature, for the 
various configurations, we introduce the partition function of the 
wire:
\begin{equation}
Z=\sum_{i}{e^{-\beta E_{i}}},\label{partifun}
\end{equation}
where $E_i$ is the energy of the $i$-th configuration and $\beta=kT$, $k$ being 
the Boltzmann constant and $T$ the temperature. The summation is performed 
over all configurations with the first cell in a given input logic state.
The probability $P_{{\rm gs}}$ of the entire system being in the 
ground state can be evaluated by taking the ratio of the Boltzmann factor for 
the ground state to the partition function:
\begin{equation}
P_{\rm gs}={\frac{e^{-\beta E_{\rm gs}}}{Z}}=\frac{1}{1+\sum_{i\ne 
{\rm gs}} {e^{{-\beta \Delta E_{i}}}}},
\label{Pgs}
\end{equation}
where $\Delta E_{i}=E_{i}-E_{{\rm gs}}$, and the sum extends over all excited
states.

As already mentioned in the introduction, $P_{{\rm gs}}$ is not the only
quantity of interest. From the point of view of applications, we are mainly 
interested in knowing the probability $P_{{\rm clo}}$ of obtaining the correct 
logic output, which is higher than $P_{{\rm gs}}$, because several 
configurations, besides the ground state, exhibit the correct polarization 
for the output cell (the cell at the end of the chain). We can compute
$P_{{\rm clo}}$ by summing over the probabilities corresponding to all such
configurations, that we label with the subscript $j$:
\begin{equation}
P_{{\rm clo}}={\frac{\sum_{j}{e^{-\beta E_{j}}}}{Z}}
\label{Pclo}
\end{equation}

We have computed both $P_{{\rm gs}}$ and $P_{{\rm clo}}$ as a function of the
ratio of the splitting $\Delta E$ between ground state and first excited 
state to $kT$. The results for a chain of 6 cells are presented in 
Fig.~\ref{vier}: in the limit  $\Delta E / (kT)\ll 1$ all the configurations
(a total of $6^{N-1}$, $N$ being the number of cells) become
equally probable and the probability $P_{{\rm gs}}$ reaches its
minimum value $1/6^{N-1}$.
The probability
of correct logic output, instead, reaches a minimum value of $1/6$, 
as a consequence of the six possible states of the output cell being 
equally probable.

It should be noted that an error probability of a few percent may appear
unacceptable for any practical circuit application, but data readout must
always be done via some detector\cite{thin}, which is characterized by a 
time constant
necessarily longer than the typical settling time of the QCA circuit. 
Therefore, each reading will be the result of an averaging procedure, and 
will be compared to a threshold value. In such a case, an error
probability of a few percent for the output state will lead in most cases
to a vanishingly small error probability for the actual output of the readout
circuit.

As already noted, the number of possible configurations for a 
circuit with $N$ cells (one of which is assumed to be the driver cell
and hence in a given, fixed configuration) is $6^{N-1}$. Thus the 
CPU time required to explore all such configurations grows exponentially
with the number of cells, which limits the length of 
the binary wires that can be investigated with this approach in a reasonable
time down to about ten cells. In order
to assess the thermal behavior of long wires, we have developed the
approximate analytical approach that will be described in the next section.

\section{Analytical model}\label{gebbo}

The development of an analytical model for the investigation of the thermal
behavior of a QCA chain requires a main simplifying assumption, in order to 
make the algebraic treatment possible: for each cell we consider only two 
configurations, the ones corresponding to the logic states 1 and 0, and,
thus, to polarization $+1$ and $-1$.
From the discussion in Sec.~\ref{enesec} it is apparent that the larger $R$,
the better this approximation is, because the role of the $X$ states is 
reduced.

Let us consider a generic 1-dimensional chain consisting of $N$ cells
and 
introduce the following 1-dimensional Ising Hamiltonian
\begin{equation}
{\cal H}=-J\,\sum_{i=1}^{N-1}\sigma_i\sigma_{i+1}\,,
\label{Hising}
\end{equation}
where for each cell labeled by the index $i$ the variable $\sigma_i$
corresponds to the polarization and therefore assumes the two
values $\pm 1$, and the positive quantity $J$ (which has 
the dimension of an energy) is related to the splitting $\Delta E$ 
between the ground state and the first excited state energies of an $N$-cell 
system  
by $J=\Delta E/2$.  Let us point out that there is a twofold degeneracy of
the ground state, corresponding to the two configurations
$\{\sigma_i=1\,,\forall i\}$ and $\{\sigma_i=-1\,,\forall i\}$. This
degeneracy is removed by enforcing the polarization state of the driver
cell, which corresponds to enforcing the configuration of one of the
boundary sites; our conventional choice is $\sigma_1=1$. In this case, 
the lowest energy state corresponds to the configuration 
$\{\sigma_i=1\,,\forall i\}$. 
The partition function of the $N$-cell system described by the
Hamiltonian~(\ref{Hising}) with the boundary condition $\sigma_1=1$ 
is given, in analogy with Eq.(\ref{partifun}), by the following expression: 
\begin{equation}
{Z}=\sum_{\{\sigma\}}e^{-\beta\,{\cal H}}\,,
\label{zeta2}
\end{equation} where
$\{\sigma\}$ stands for the summation over all possible states,  
i.e. $\{\sigma_1=1\,,\sigma_i=\pm 1\,,i=2\dots N\}$. 
This last expression can be written as 
\begin{equation}
{Z}=\sum_{\sigma_2\dots\sigma_N}V(1,\sigma_2)\,V(\sigma_2,\sigma_3)\dots
V(\sigma_{N-1},\sigma_N)\,,
\label{vtmp}
\end{equation}
where $V(\sigma,\sigma^{\prime})=e^{\beta\sigma\sigma^{\prime}}$. In
order to compute the r.h.s. of Eq.~(\ref{vtmp}), the usual procedure consists
in introducing the transfer matrix~\cite{stat}
\begin{equation}
{\cal V}=\left(\begin{array}{cc} e^{\beta J} & e^{-\beta J} \\
e^{-\beta J} & e^{\beta J} \end{array} \right ), 
\label{tram}
\end{equation}
whose eigenvalues are
\begin{equation}
\lambda_+=e^{\beta J} + e^{-\beta J}\,,\quad\quad 
\lambda_-=e^{\beta J} - e^{-\beta J}\,.
\end{equation}
The expression for the matrix ${\cal V}^{N-1}$ is given by 
\begin{equation}
{\cal V}^{N-1}=\left(\begin{array}{cc}
\frac{\lambda_+^{N-1}+\lambda_-^{N-1}}{2} &
\frac{\lambda_+^{N-1}-\lambda_-^{N-1}}{2} \\
\frac{\lambda_+^{N-1}-\lambda_-^{N-1}}{2} &
\frac{\lambda_+^{N-1}+\lambda_-^{N-1}}{2}
\end{array} \right ).
\end{equation}
It then follows that the partition function~(\ref{vtmp}) reads
\begin{equation}
{Z}=[{\cal V}^{N-1}]_{11}+[{\cal V}^{N-1}]_{12}=
(e^{\beta J}+e^{-\beta J})^{N-1},
\label{zetaising}
\end{equation}
where the subscripts indicate specific elements of the ${\cal V}^{N-1}$ matrix. 
This explicit formula for the partition function allows us to derive
an analytical expression for the probability of the system being in
its ground state as a function of the temperature and of the energy
splitting between the two lowest states. Since the ground state energy
for the Hamiltonian~(\ref{Hising}) is $E_{\rm gs}=-J(N-1)$, we obtain

\begin{equation}
P_{\rm gs}=\frac{e^{-\beta E_{\rm gs}}}{{Z}}=
\frac{e^{\beta J(N-1)}}{{Z}}=\frac{1}{(1+e^{-\beta\Delta E})^{N-1}}.
\label{prob2}
\end{equation}

Finally, we can derive an analytical expression also for the probability of
obtaining the correct logic output, in analogy with what we have already done 
in the numerical case.
We need to determine the occupation probability of a generic state with 
$\sigma_1=\sigma_N=1$, which corresponds to having the correct output,
because the polarization of the $N-$th cell (output cell) is the same
as that of the first cell.
To this purpose, we evaluate
the following  ``reduced'' partition function: 
\begin{equation}
{Z}_R=\sum_{\sigma_2\dots\sigma_{N-1}}V(1,\sigma_2)\,
V(\sigma_2,\sigma_1)\dots V(\sigma_{N-1},1)\,,
\label{repa}
\end{equation} 
where again
$V(\sigma,\sigma^{\prime})=e^{\beta\sigma\sigma^{\prime}}$. Using the
transfer matrix~(\ref{tram}), it follows that 
${Z}_R=[{\cal V}^{N-1}]_{11}$, and hence 
\begin{equation}
P_{\rm clo}=\frac{{Z}_R}{{Z}}=
\frac{[{\cal V}^{N-1}]_{11}}
{[{\cal V}^{N-1}]_{11}+[{\cal V}^{N-1}]_{12}}=
\frac{1}{2}\,
\left[1+\left(\tanh(\beta\Delta E/2)\right)^{N-1}\right]
\label{porod2}
\end{equation}

The above derived analytical expressions have been used to compute $P_{\rm gs}$
and $P_{\rm clo}$ as a function of temperature for a chain of 6 cells,
cell size $a=40$~nm, cell separation $d=100$~nm. Results are presented 
with dashed lines in Fig.~\ref{fuenf}, together with those obtained with the 
numerical technique (solid lines). For temperatures below about 2~K (those
for which reasonably low error probabilities can be achieved) the
analytical model provides values that are in almost perfect agreement with
those from the more detailed numerical approach. The situation differs at
higher temperatures, because higher energy configurations, containing cells
in $X$ states, start being occupied and are properly handled by the numerical
model, while they are not at all included in the analytical approach. In 
particular,
while for large values of the temperature the numerical $P_{\rm clo}$ tends 
to $1/6$, as
previously discussed, the analytical $P_{\rm clo}$ approaches the value $1/2$,
because the output cell can be in one of two states with the same probability.
Analogous considerations can be made for $P_{\rm gs}$, which becomes extremely
small ($1/6^5$) for higher temperatures in the numerical case, while drops
just to $1/2^5$ in the analytical case, since there are $2^5$ possible 
configurations. 

In Fig.~\ref{sechs}, $P_{\rm clo}$ and $P_{\rm gs}$ are reported as a function 
of the ratio $\Delta E/(kT)$ (with a semilogarithmic scale) for the analytical 
model
(thick solid line), and for the numerical model with $R=2$ (thin solid line),
$R=2.5$ (dashed line), and $R=4$ (dotted line). As expected, the agreement 
improves with increasing $R$, because of the reduced relevance of the $X$
states. For $\Delta E/(kT)$ of the order of a few units, the error 
probability becomes very small and the analytical expression can be reliably
used to evaluate it. 

In particular, the analytical expression allows us to provide estimates of
the maximum operating temperature for a QCA chain formed by a given number of
cells. 
We have computed the maximum operating temperature 
allowing a given correct logic output probability, as
a function of the number of cells: results are reported in Fig.~\ref{sieben} 
for $P_{\rm clo}=0.6$ (solid line), $0.9$ (dashed line), $0.99$ (dotted line) 
and cell size $a=40$~nm, intercell separation $d=100$~nm.
The maximum operating temperature, for a number of cells above a few tens,
drops logarithmically, which leads to a linear behavior in the logarithmic
representation of Fig.~\ref{sieben}. 

\section{Conclusions}\label{conclu}

We have developed both a numerical and an analytical approach to the 
investigation of the thermal dependence of QCA wire operation. Both
methods are based on a semiclassical approach, in which electrons are 
considered as classical particles interacting via the Coulomb force,
with, however, the possibility of tunneling between the quantum dots belonging 
to the same cell. The electrostatic energy associated with each configuration
has been evaluated and used for the calculation of the occupancies, via
the partition function. 

Numerical results have been derived for wires
with six-state cells, which realistically reproduce the behavior of
QCA systems, provided that the confinement in each quantum dot is strong
enough. The numerical procedure thus developed is general and is currently 
being applied to the investigation of thermal limitations for simple
logic gates, including the effect of spurious, $X$ states. 

The analytical approach has allowed a detailed analysis of the error
probability due to thermal excitations in arbitrarily long wires, generalizing
the findings of previous studies, and the possibility of extending it to 
selected basic gates is being investigated. 

It is clear from our results that the operating temperature depends 
on the ratio of the energy splitting $\Delta E$ to $kT$. It could therefore
be raised by increasing $\Delta E$, which means reducing the dielectric 
permittivity or scaling down cell dimensions. As already mentioned, the
silicon-on-insulator material system offers better perspectives of 
higher-temperature operation, due to the lower permittivity of silicon 
oxide. However, scaling down in any semiconductor
implementation is limited by the increasing precision
requirements\cite{JAPnos}, therefore a trade-off between manufacturability and
operating temperature has to be accepted. Implementations at the molecular
level could provide better opportunities, due to the reduced dimensions, but 
their actual feasibility is still being assessed.

\acknowledgments
This work has been supported by the ESPRIT Project N.~28667 ANSWERS,
Autonomous Nanoelectronic Systems With Extended Replication and Signalling.

\begin{figure}
\caption {a) Chain of 6 QCA cells, with the indication of the cell
size $a$ and of the intercell separation $d$; b) representation of the
six possible states of a cell with two electrons.}
\label{eins}
\end{figure}

\begin{figure}
\caption{Energy spectrum and cell configurations for a six-cell wire, for 
two values of the ratio $R$ of intercell separation $d$ to cell size $a$,
with $a=40$~nm.} 
\label{zwei}
\end{figure}

\begin{figure}
\caption{Energy split between the first excited state and the ground state
for a chain of 2 (dotted line), 3 (dashed line), 6 (solid line) cells as
a function of the ratio $R$ of the intercell separation $d$ to the cell
size $a$, assumed to be 40~nm. The energy split as a function of the number
of cells is reported in the inset, for a value $R=2.5$.}
\label{drei}
\end{figure}

\begin{figure}
\caption{Plot of the probability of correct logic output $P_{\rm clo}$ and
of the probability of ground state occupancy $P_{\rm gs}$ for a chain of six 
cells, with R=2.5, as a function of the ratio $\Delta E/(kT)$}
\label{vier}
\end{figure}

\begin{figure}
\caption{Plot of the probability of correct logic output $P_{\rm clo}$ and
of the probability of ground state occupancy $P_{\rm gs}$ for a chain of six 
cells as a function of temperature, computed with the numerical six-state
model (solid lines) and with the analytical two-state model (dashed lines).
The cell size $a$ is 40~nm and the intercell separation $d$ is 100~nm.}
\label{fuenf}
\end{figure}

\begin{figure}
\caption{Plot of $P_{\rm clo}$ and $P_{\rm gs}$ as a function of $\Delta E/(kT)$
for the analytical two-state model (thick solid line) and for the 
numerical six-state model with $R=2$ (thin solid line), $R=2.5$ (dashed line),
and $R=4$ (dotted line).} 
\label{sechs}
\end{figure}

\begin{figure}
\caption{Maximum operating temperature of a QCA binary wire as a function of 
the number of cells, for obtaining a probability of correct logic output
$0.6$ (solid line), $0.9$ (dashed line), and $0.99$ (dotted line). The
cell size is $40$~nm and the intercell separation is $100$~nm.}
\label{sieben}
\end{figure}

\end{document}